\documentclass[conference]{IEEEtran}
\IEEEoverridecommandlockouts
\usepackage{cite}
\usepackage{amsmath,amssymb,amsfonts, bm}
\usepackage{algorithmic}
\usepackage{graphicx}
\usepackage{textcomp}
\usepackage{xcolor}
\usepackage{dirtytalk}
\usepackage{multirow}
\usepackage{array}
\usepackage{colortbl}
\usepackage{soul}
\usepackage{stmaryrd}
\usepackage{flushend}
\def\BibTeX{{\rm B\kern-.05em{\sc i\kern-.025em b}\kern-.08em
    T\kern-.1667em\lower.7ex\hbox{E}\kern-.125emX}}

\usepackage{fancyhdr}
\fancypagestyle{firstpage}{
  \fancyhf{}
  \fancyhead[C]{\small Extended abstract for a poster at the \textit{IEEE Communications Theory Workshop (CTW)}, Banff, Canada, May 2020. (Moved to June  2021.)}
}

\begin{document}

\title{Error Resilient Collaborative Intelligence via Low-Rank Tensor Completion 
}

\author{\IEEEauthorblockN{Lior Bragilevsky}
\IEEEauthorblockA{\textit{School of Engineering Science} \\
\textit{Simon Fraser University}\\
Burnaby, BC, Canada\\
lbragile@sfu.ca}
\and
\IEEEauthorblockN{Ivan V. Baji\'{c}}
\IEEEauthorblockA{\textit{School of Engineering Science} \\
\textit{Simon Fraser University}\\
Burnaby, BC, Canada\\
ibajic@ensc.sfu.ca}
}

\maketitle

\begin{abstract}
In the race to bring Artificial Intelligence (AI) to the edge, \textit{collaborative intelligence} has emerged as a promising way to lighten the computation load on edge devices that run applications based on Deep Neural Networks (DNNs). Typically, a deep model is split at a certain layer into edge and cloud sub-models. The deep feature tensor produced by the edge sub-model is transmitted to the cloud, where the remaining computationally intensive workload is performed by the cloud sub-model. The communication channel between the edge and cloud is imperfect, which will result in missing data in the deep feature tensor received at the cloud side. In this study, we examine the effectiveness of four low-rank tensor completion methods in recovering missing data in the deep feature tensor. We consider both sparse tensors, such as those produced by the VGG16 model, as well as non-sparse tensors, such as those produced by ResNet34 model. We study tensor completion effectiveness in both conplexity-constrained and unconstrained scenario.
\end{abstract}


\thispagestyle{firstpage}

\section{Introduction}
Collaborative Intelligence (CI)~\cite{CI_ICASSP_2021} 
is an AI deployment strategy that leverages both edge-based and cloud-based resources to make DNN computing faster and more efficient. In CI, a deep model is split into an edge sub-model and a cloud sub-model. For example, an edge sub-model may consist of the initial $m$ layers of a DNN, while the cloud sub-model is made up of the remaining DNN layers. When an input signal is captured by an edge sensor, the edge sub-model processes the signal and produces a tensor of deep features, which is then transmitted to the cloud for subsequent processing. Due to the imperfect channel between the edge and the cloud, tensor data may be damaged or missing. Hence, error control  must be deployed to achieve seamless operation of a CI system. 

We study four methods for recovery of missing data in a deep feature tensor: Simple Low Rank Tensor Completion (SiLRTC)~\cite{main_algorithms}, High Accuracy Low Rank Tensor Completion (HaLRTC)~\cite{main_algorithms}, Fused Canonical Polyadic (FCP) decomposition~\cite{fcp_algorithm}, and Adaptive Linear Tensor Completion (ALTeC)~\cite{ALTeC}. All these methods are based on the low-rank tensor assumption.\footnote{There are more recent methods based on different modeling of feature tensors, e.g.~\cite{IVB_ICC_2021, Hans_ICIP_2021}, but these are outside the scope of this comparison.} SiLRTC and HaLRTC are general methods applicable to any kind of tensor; FCP has two versions, one for sparse and the other for non-sparse tensors, whereas ALTeC is trained specifically for the model whose tensors it is meant to complete. Further details are given in the respective references. 

\section{Experiments}
\subsection{Setup}
We assume deep feature tensors are packetized row-by-row, similarly to rows of macroblocks in video streaming~\cite{Wang_etal_2002}. We use two image classification models in the experiments: VGG16~\cite{vgg} (which produces sparse tensors) and ResNet34~\cite{resnet} (which produces non-sparse tensors). VGG16 was split at layer \say{block4\_pool}, such that the edge sub-model had 5.52\% of the model's total number of parameters. ResNet34 was split at layer \say{add\_7}, such that the edge sub-model had 6.19\% of the total number of parameters. Test data is a randomly-selected subset of 1,000 images from the ILSVRC~\cite{ilsvrc} validation set, which were different from the 5,000 images on which ALTeC was trained. We consider an independent random packet loss with probability $p_{loss} \in \{5\%, 10\%, 15\%, 20\%,25\%, 30\%\}$. It is assumed that at the receiver (cloud sub-model), missing packets are identified via packet sequence numbers provided by a transport-layer protocol such as the Real-time Transport Protocol (RTP)~\cite{RTP}. For each test image, 100 realizations of packet loss were simulated.

\begin{figure*}
    \centering
    \vspace{-12pt}
    \includegraphics[width=0.93\textwidth]{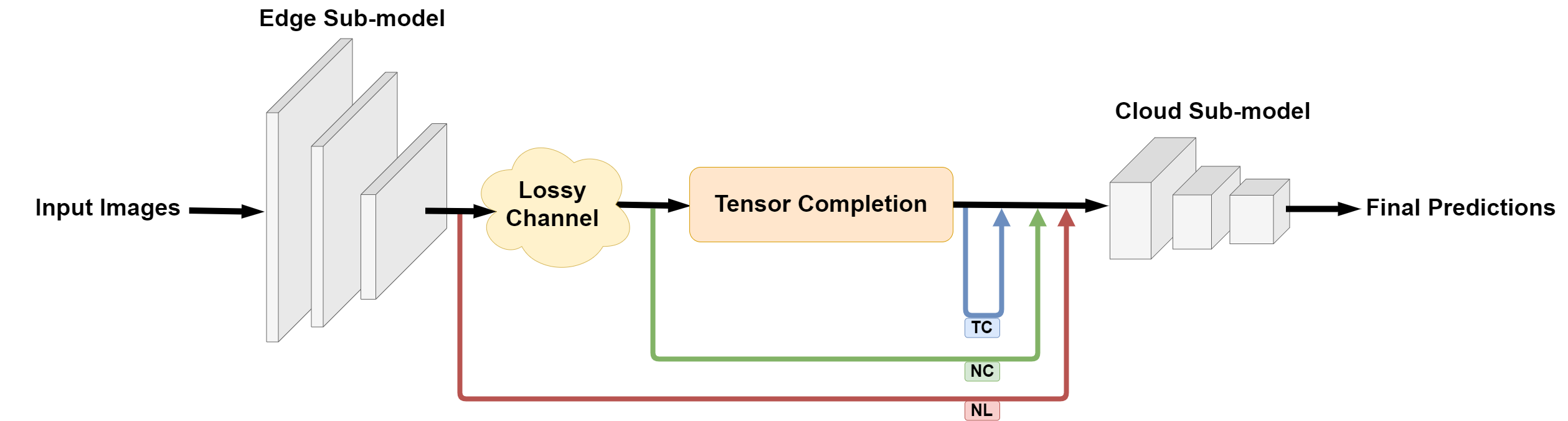}
    \caption{Test conditions: NL (no loss, baseline), NC (no tensor completion), TC (tensor completion)}
    \label{fig:pred_cases}
    \vspace{-0.2cm}
\end{figure*}

Since we focus on image classification models, we measure classification accuracy under three conditions (Fig.~\ref{fig:pred_cases}): (1) no loss (NL); (2) no tensor completion (NC), where all missing data is assumed to be zero; (3) tensor completion (TC), where a specific tensor completion algorithm is performed on the corrupt tensor.
The average Top-1 classification accuracy for the three cases ($\mu_{\text{NL}},\ \mu_{\text{NC}},$ and $\mu_{\text{TC}}$) and the standard deviation of Top-1 classification accuracy under NC and TC conditions ($\sigma_{\text{NC}}$ and $\sigma_{\text{TC}}$) are measured. All tensor completion algorithms are tested on a Linux machine running Ubuntu 16.04.5 LTS, with Intel Core i7-6800K CPU @ 3.40GHz, 128GB RAM, Titan X GPU with 12GB memory, with Tensorflow 1.9.0 and Keras 2.2.4. For each method, the ``default'' result is when the method is run until convergence, whereas ``speed-matched'' result is obtained when the speed is matched to that of ALTeC, the fastest of the four methods.

\subsection{Results}
On sparse tensors from VGG16 (Table~\ref{tab:results}), all methods offer similar performance. There is no clear winner at the 95\% significance level, as confirmed by Welch's t-test for samples with unequal variance. On the non-sparse tensors from ResNet34 (Table~\ref{tab:results-resnet}), HaLRTC has the best performance in the ``default'' setting for loss above 5\%, whereas ALTeC is the best in the ``speed-matched'' setting across all loss rates.

\begin{table}[t!]
\vspace{-3pt}
\caption{VGG16 classification accuracy results. No method outperforms all others at the 95\% significance level.}
\vspace{-3pt}
\label{tab:results}
\setlength{\tabcolsep}{3pt}
\setlength\extrarowheight{3pt}
\scalebox{0.85}{
\begin{tabular}{|>{\centering\arraybackslash}p{0.7cm}|>{\centering\arraybackslash}p{1.25cm}|>{\centering\arraybackslash}p{0.85cm}|>{\centering\arraybackslash}p{0.85cm}|>{\centering\arraybackslash}p{0.7cm}|}
\multicolumn{5}{c}{}\vspace{0.5pt}\\
\hline
$\bm{{p}_{loss}}$ & \textbf{Algorithm} & $\bm{\mu_{\text{NL}}}$ & $\bm{\mu_{\text{NC}}}$ & $\bm{\sigma_{\text{NC}}}$ \\
\hline\hline
\multirow{4}{*}{$5\%$}  & SiLRTC &56.20\%  &55.96\%   &0.41\%   \\\cline{2-5}
                        & HaLRTC &56.20\%  &55.96\%   &0.41\%  \\\cline{2-5}
                        & FCP    &56.20\%  &55.96\%   &0.41\%  \\\cline{2-5}
                        & ALTeC  &56.20\%  &55.96\%   &0.41\%   \\\hline\hline
\multirow{4}{*}{$10\%$} & SiLRTC &56.20\%  &55.50\%   &0.55\%  \\\cline{2-5}
                        & HaLRTC &56.20\%  &55.50\%   &0.55\%  \\\cline{2-5}
                        & FCP    &56.20\%  &55.50\%   &0.55\%  \\\cline{2-5}
                        & ALTeC  &56.20\%  &55.50\%   &0.55\%   \\\hline\hline
\multirow{4}{*}{$15\%$} & SiLRTC &56.20\%  &54.76\%   &0.60\%   \\\cline{2-5}
                        & HaLRTC &56.20\%  &54.76\%   &0.60\%  \\\cline{2-5}
                        & FCP    &56.20\%  &54.76\%   &0.60\%  \\\cline{2-5}
                        & ALTeC  &56.20\%  &54.76\%   &0.60\%   \\\hline\hline
\multirow{4}{*}{$20\%$} & SiLRTC &56.20\%  &54.18\%   &0.63\%   \\\cline{2-5}
                        & HaLRTC &56.20\%  &54.18\%   &0.63\%   \\\cline{2-5}
                        & FCP    &56.20\%  &54.18\%   &0.63\%   \\\cline{2-5}
                        & ALTeC  &56.20\%  &54.18\%   &0.63\%   \\\hline\hline
\multirow{4}{*}{$25\%$} & SiLRTC &56.20\%  &53.45\%   &0.79\%   \\\cline{2-5}
                        & HaLRTC &56.20\%  &53.45\%   &0.79\%   \\\cline{2-5}
                        & FCP    &56.20\%  &53.45\%   &0.79\%   \\\cline{2-5}
                        & ALTeC  &56.20\%  &53.45\%   &0.79\%   \\\hline\hline
\multirow{4}{*}{$30\%$} & SiLRTC &56.20\%  &52.57\%   &0.77\%   \\\cline{2-5}
                        & HaLRTC &56.20\%  &52.57\%   &0.77\%   \\\cline{2-5}
                        & FCP    &56.20\%  &52.57\%   &0.77\%   \\\cline{2-5}
                        & ALTeC  &56.20\%  &52.57\%   &0.77\%   \\\hline
\end{tabular}
\hspace{0.0005\textwidth}
\begin{tabular}{|>{\centering\arraybackslash}p{0.85cm}|>{\centering\arraybackslash}p{0.7cm}|}
\hline
\multicolumn{2}{|c|}{\textbf{Default}}\\
\hline
$\bm{\mu_{\text{TC}}}$ & $\bm{\sigma_{\text{TC}}}$\\\hline\hline
    56.09\%   &0.39\%\\\cline{1-2}
    56.06\%   &0.36\%\\\cline{1-2}
    56.09\%   &0.41\%\\\cline{1-2}
    56.07\%   &0.40\%\\\hline\hline
    55.67\%   &0.44\% \\\cline{1-2}
    55.75\%   &0.37\%\\\cline{1-2}
    55.78\%   &0.51\%\\\cline{1-2}
    55.70\%   &0.48\%\\\hline\hline
    54.99\%   &0.58\%\\\cline{1-2}
    55.14\%   &0.44\%\\\cline{1-2}
    55.17\%   &0.55\%\\\cline{1-2}
    55.11\%   &0.56\%\\\hline\hline
    54.51\%   &0.61\%\\\cline{1-2}
    54.67\%   &0.55\%\\\cline{1-2}
    54.74\%   &0.65\%\\\cline{1-2}
    54.64\%   &0.63\%\\\hline\hline
    53.95\%   &0.69\%\\\cline{1-2}
    54.19\%   &0.67\%\\\cline{1-2}
    54.16\%   &0.71\%\\\cline{1-2}
    54.03\%   &0.75\%\\\hline\hline
    53.13\%   &0.73\%\\\cline{1-2}
    53.39\%   &0.67\%\\\cline{1-2}
    53.31\%   &0.78\%\\\cline{1-2}
    53.25\%   &0.81\%\\\hline
\end{tabular}
\hspace{0.0005\textwidth}
\begin{tabular}{|>{\centering\arraybackslash}p{0.85cm}|>{\centering\arraybackslash}p{0.7cm}|}
\hline
\multicolumn{2}{|c|}{\textbf{Speed-matched}}\\
\hline
$\bm{\mu_{\text{TC}}}$ & $\bm{\sigma_{\text{TC}}}$\\\hline\hline
    55.96\%   &0.41\% \\\cline{1-2}
    55.96\%   &0.41\%\\\cline{1-2}
    55.99\%   &0.44\%\\\cline{1-2}
    56.07\%   &0.40\%\\\hline\hline
    55.53\%   &0.52\% \\\cline{1-2}
    55.51\%   &0.54\%\\\cline{1-2}
    55.78\%   &0.53\%\\\cline{1-2}
    55.70\%   &0.48\%\\\hline\hline
    54.79\%   &0.62\%\\\cline{1-2}
    54.75\%   &0.59\%\\\cline{1-2}
    55.15\%   &0.59\%\\\cline{1-2}
    55.11\%   &0.56\%\\\hline\hline
    54.24\%   &0.64\%\\\cline{1-2}
    54.21\%   &0.64\%\\\cline{1-2}
    54.72\%   &0.67\%\\\cline{1-2}
    54.64\%   &0.63\%\\\hline\hline
    53.51\%   &0.76\%\\\cline{1-2}
    53.48\%   &0.80\%\\\cline{1-2}
    54.16\%   &0.72\%\\\cline{1-2}
    54.03\%   &0.75\%\\\hline\hline
    52.69\%   &0.78\%\\\cline{1-2}
    52.65\%   &0.78\%\\\cline{1-2}
    53.26\%   &0.74\%\\\cline{1-2}
    53.25\%   &0.81\%\\\hline
\end{tabular}}
\end{table}

\begin{table}[t!]
\vspace{-3pt}
\caption{ResNet34 classification accuracy results. Best performance at the 95\% significance level is indicated in bold.}
\vspace{-3pt}
\label{tab:results-resnet}
\setlength{\tabcolsep}{3pt}
\setlength\extrarowheight{3pt}
\scalebox{0.85}{
\begin{tabular}{|>{\centering\arraybackslash}p{0.7cm}|>{\centering\arraybackslash}p{1.25cm}|>{\centering\arraybackslash}p{0.85cm}|>{\centering\arraybackslash}p{0.85cm}|>{\centering\arraybackslash}p{0.7cm}|}
\multicolumn{5}{c}{}\vspace{0.5pt}\\
\hline
$\bm{p}_{loss}$ & \textbf{Algorithm} & $\bm{\mu_{\text{NL}}}$ & $\bm{\mu_{\text{NC}}}$ & $\bm{\sigma_{\text{NC}}}$ \\
\hline\hline
\multirow{4}{*}{$5\%$}  & SiLRTC &58.10\%  &57.57\%   &0.61\%   \\\cline{2-5}
                        & HaLRTC &58.10\%  &57.57\%   &0.61\%  \\\cline{2-5}
                        & FCP    &58.10\%  &57.57\%   &0.61\%   \\\cline{2-5}
                        & ALTeC  &58.10\%  &57.57\%   &0.61\%  \\\hline\hline
\multirow{4}{*}{$10\%$} & SiLRTC &58.10\%  &54.57\%   &0.68\%  \\\cline{2-5}
                        & HaLRTC &58.10\%  &54.57\%   &0.68\%  \\\cline{2-5}
                        & FCP    &58.10\%  &54.57\%   &0.68\%   \\\cline{2-5}
                        & ALTeC  &58.10\%  &54.57\%   &0.68\%  \\\hline\hline
\multirow{4}{*}{$15\%$} & SiLRTC &58.10\%  &49.30\%   &0.78\%   \\\cline{2-5}
                        & HaLRTC &58.10\%  &49.30\%   &0.78\%  \\\cline{2-5}
                        & FCP    &58.10\%  &49.30\%   &0.78\%   \\\cline{2-5}
                        & ALTeC  &58.10\%  &49.30\%   &0.78\%  \\\hline\hline
\multirow{4}{*}{$20\%$} & SiLRTC &58.10\%  &40.87\%   &0.86\%   \\\cline{2-5}
                        & HaLRTC &58.10\%  &40.87\%   &0.86\%   \\\cline{2-5}
                        & FCP    &58.10\%  &40.87\%   &0.86\%   \\\cline{2-5}
                        & ALTeC  &58.10\%  &40.87\%   &0.86\%   \\\hline\hline
\multirow{4}{*}{$25\%$} & SiLRTC &58.10\%  &29.11\%   &0.86\%   \\\cline{2-5}
                        & HaLRTC &58.10\%  &29.11\%   &0.86\%   \\\cline{2-5}
                        & FCP    &58.10\%  &29.11\%   &0.86\%   \\\cline{2-5}
                        & ALTeC  &58.10\%  &29.11\%   &0.86\%   \\\hline\hline
\multirow{4}{*}{$30\%$} & SiLRTC &58.10\%  &15.72\%   &0.77\%   \\\cline{2-5}
                        & HaLRTC &58.10\%  &15.72\%   &0.77\%   \\\cline{2-5}
                        & FCP    &58.10\%  &15.72\%   &0.77\%   \\\cline{2-5}
                        & ALTeC  &58.10\%  &15.72\%   &0.77\%   \\\hline
\end{tabular}
\hspace{0.0005\textwidth}
\begin{tabular}{|>{\centering\arraybackslash}p{0.85cm}|>{\centering\arraybackslash}p{0.7cm}|}
\hline
\multicolumn{2}{|c|}{\textbf{Default}}\\
\hline
$\bm{\mu_{\text{TC}}}$ & $\bm{\sigma_{\text{TC}}}$\\\hline\hline
    57.77\%   &0.49\%\\\cline{1-2}
    57.94\%   &0.37\%\\\cline{1-2}
    57.92\%   &0.43\%\\\cline{1-2}
    58.04\%   &0.44\% \\\hline\hline
    56.47\%   &0.60\%\\\cline{1-2}
    \textbf{57.65}\%   &0.46\%\\\cline{1-2}
    56.56\%   &0.66\%\\\cline{1-2}
    57.18\%   &0.61\%\\\hline\hline
    53.89\%   &0.64\%\\\cline{1-2}
    \textbf{57.02}\%   &0.51\%\\\cline{1-2}
    53.96\%   &0.75\%\\\cline{1-2}
    55.09\%   &0.71\%\\\hline\hline
    49.61\%   &0.77\%\\\cline{1-2}
    \textbf{56.26}\%   &0.60\%\\\cline{1-2}
    49.76\%   &0.76\%\\\cline{1-2}
    51.99\%   &0.72\%\\\hline\hline
    43.56\%   &0.87\%\\\cline{1-2}
    \textbf{55.09}\%   &0.65\%\\\cline{1-2}
    44.10\%   &0.81\%\\\cline{1-2}
    47.52\%   &0.67\%\\\hline\hline
    34.56\%   &0.89\%\\\cline{1-2}
    \textbf{53.63}\%   &0.68\%\\\cline{1-2}
    36.06\%   &0.76\%\\\cline{1-2}
    41.23\%   &0.80\%\\\hline
\end{tabular}
\hspace{0.0005\textwidth}
\begin{tabular}{|>{\centering\arraybackslash}p{0.85cm}|>{\centering\arraybackslash}p{0.7cm}|}
\hline
\multicolumn{2}{|c|}{\textbf{Speed-matched}}\\
\hline
$\bm{\mu_{\text{TC}}}$ & $\bm{\sigma_{\text{TC}}}$\\\hline\hline
    57.75\%   &0.53\% \\\cline{1-2}
    57.75\%   &0.61\%\\\cline{1-2}
    57.59\%   & 0.60\%\\\cline{1-2}
    \textbf{58.04}\%   &0.44\%\\\hline\hline
    56.12\%   &0.66\% \\\cline{1-2}
    54.57\%   &0.68\%\\\cline{1-2}
    55.98\%   &0.69\%\\\cline{1-2}
    \textbf{57.18}\%   &0.61\%\\\hline\hline
    52.84\%   &0.71\%\\\cline{1-2}
    49.31\%   &0.78\%\\\cline{1-2}
    53.20\%   &0.78\%\\\cline{1-2}
    \textbf{55.09}\%   &0.71\%\\\hline\hline
    48.64\%   &0.80\%\\\cline{1-2}
    40.87\%   &0.86\%\\\cline{1-2}
    49.10\%   &0.87\%\\\cline{1-2}
    \textbf{51.99}\%   &0.72\%\\\hline\hline
    41.40\%   &0.99\%\\\cline{1-2}
    29.11\%   &0.87\%\\\cline{1-2}
    43.07\%   &0.82\%\\\cline{1-2}
    \textbf{47.52}\%   &0.67\%\\\hline\hline
    31.85\%   &0.83\%\\\cline{1-2}
    15.73\%   &0.77\%\\\cline{1-2}
    34.93\%   &0.80\%\\\cline{1-2}
    \textbf{41.23}\%   &0.80\%\\\hline
\end{tabular}}
\end{table}


\begin{thebibliography}{00}

\bibitem{CI_ICASSP_2021} I. V. Baji\'{c}, W. Lin, and Y. Tian,  \say{Collaborative intelligence: Challenges and opportunities,} \textit{Proc. IEEE ICASSP}, Jun. 2021. To appear.




\bibitem{main_algorithms} J. Liu, P. Musialski, P. Wonka, and J. Ye, \say{Tensor completion for estimating missing values in visual data,} \textit{IEEE Trans. Pattern Analysis and Machine Intelligence}, vol. 35, no. 1, pp. 208-220, Jan. 2012.

\bibitem{fcp_algorithm} Y. Wu, H. Tan, Y. Li, J. Zhang, X. Chen, \say{A fused CP factorization method for incomplete tensors,} \textit{IEEE Transactions on Neural Networks and Learning Systems}, vol. 30, no. 3, pp. 751-764, Mar. 2019.

\bibitem{ALTeC} L. Bragilevsky and I. V. Baji\'{c}, \say{Tensor completion methods for collaborative intelligence,} \textit{IEEE Access}, vol. 8, pp. 41162-41174, Feb. 2020.

\bibitem{IVB_ICC_2021} I. V. Baji\'{c}, \say{Latent space inpainting for loss-resilient collaborative object detection,} \textit{Proc. IEEE ICC}, Jun. 2021. To appear.

\bibitem{Hans_ICIP_2021} A. Dhondea, R. A. Cohen, and I. V. Baji\'{c}, \say{CALTeC: Content-adaptive linear tensor completion for collaborative intelligence,} \textit{Proc. IEEE ICIP}, Sep. 2021. To appear.

\bibitem{Wang_etal_2002} Y. Wang, J. Ostermann, and Y.-Q. Zhang, \textit{Video Processing and Communications}, Prentice-Hall, 2002.

\bibitem{vgg} K. Simonyan and A. Zisserman, \say{Very deep convolutional networks for large-scale image recognition,} \textit{Proc. ICLR'15}, May 2015.

\bibitem{resnet} K. He, X. Zhang, S. Ren and J. Sun, \say{Deep residual learning for image recognition,} \textit{Proc. IEEE CVPR'16}, pp. 770-778, Jun. 2016.

\bibitem{ilsvrc} O. Russakovsky et al., 
\say{ImageNet large scale visual recognition challenge,} \textit{Int. J. Comp. Vis.}, vol. 115, no. 3, pp. 211-252, Dec. 2015.


\bibitem{RTP} H. Schulzrinne, S.  Casner, R. Frederick, and V. Jacobson, \say{RTP: A transport protocol for real-time applications,} RFC 3550, Jul. 2003.











\end{thebibliography}
\end{document}